\documentclass[pra,twocolumn,showpacs]{revtex4}%
\usepackage{amsfonts}
\usepackage{amsmath}
\usepackage{amssymb}
\usepackage{graphicx}
\usepackage{subfigure}%
\begin{document}
\title{The entanglement dynamics of interacting qubits embedded in a spin environment
with Dzyaloshinsky-Moriya term}
\author{Wen-Long You}
\author{Yu-Li Dong}
\altaffiliation{Email: yldong@suda.edu.cn}

\affiliation{School of Physical Science and Technology, Suzhou
University, Suzhou, Jiangsu 215006, People's Republic of China}

\begin{abstract}
We investigate the entanglement dynamics of two interacting qubits
in a spin environment, which is described by an XY model with
Dzyaloshinsky-Moriya (DM) interaction. The competing effects of
environmental noise and interqubit coupling on entanglement
generation for various system parameters are studied. We find that
the entanglement generation is suppressed remarkably in
weak-coupling region at quantum critical point (QCP). However, the
suppression of the entanglement generation at QCP can be compensated
both by increasing the DM interaction and by decreasing the
anisotropy of the spin chain. Beyond the weak-coupling region, there
exist resonance peaks of concurrence when the system-bath coupling
equals to external magnetic field. We attribute the presence of
resonance peaks to the flat band of the self-Hamiltonian. These
peaks are highly sensitive to anisotropy parameter and DM
interaction.

\end{abstract}

 \maketitle

\section{Introduction}

\label{introduction} Since the discovery of quantum mechanics,
quantum entanglement has played an important role in quantum
information processing (QIP), such as superdense coding \cite{ben},
teleportation \cite{ben92} and quantum algorithms \cite{joz}. In
recent years, there has been a growing interest in entanglement
dynamics by using the coherent manipulation in solid state systems.
One of the most natural candidates is spin chain, which can be
simulated by means of cold atoms in optical lattices \cite{duan03}
or by coupled microcavities \cite{har07}, and has been extensively
studied in numerous works \cite{os02,ost02,gun01,ar01,zhu}.
Especially much attention has been paid to the relation between
entanglement and quantum phase transitions (QPTs) which are driven
purely by quantum fluctuation occurring at zero temperature
\cite{wang,Connor,Gu}. The QPT is related to a dramatic change in
the ground-state properties of the system as an external
non-temperature parameter varies across the transition point.
Consequently, the dramatic change in the structure of the ground
state should result in a great difference between the quantum
correlation on both sides of the quantum critical point (QCP).
Naturally, the entanglement should be able to characterize the
change. For example, Osterloh \textit{et al.} \cite{ost02} has
proven that the derivative of the nearest-neighbor entanglement
diverges at QCP. Due to the dynamical ultrasensitivity of the
induced quantum critical system \cite{quan}, quantum entanglement
can be treated as a tool \cite{oru,amico} to characterize QPTs.

It's worth noting that the systems considered are closed, i.e.,
isolated systems have no interaction with their external
environment. However, real physical systems are never isolated,
since the coupling between system and the surrounding environment is
inevitable. The quantum dynamics of physical systems is always
complicated by their coupling to many 'environmental' modes. The
dominant environmental effects are localized modes at low
temperature, which are usually described by spin-bath model. An
interesting phenomenon is that the coupling process between system
and bath shows a duality of influence on quantum system. On one
hand, the coupling can assist people to achieve some tasks in QIP
\cite{ple,bos}. One of the focuses is the induced entanglement
between the two noninteracting qubits. For instance, Yi \textit{et
al.} \cite{yi} showed that the entanglement changed dramatically
along the line of critical points of spin bath. Subsequently, an
enhanced effect of induced entanglement near the critical point was
demonstrated \cite{sun}. These investigations exhibit a new
perspective to engineer protocols for entanglement generation. On
the other hand, the coupling between system and bath can play a role
as decoherence. It can transfer a pure ensemble of qubits to a
mixture of classical ones \cite{zur} and lead to asymptotical
disappearance of system entanglement. In some cases, the
entanglement will vanish even in finite time \cite{yu}. Great
efforts have been devoted to the study of the decoherence caused by
the spin bath \cite{bose04,zur05}. One of the most common models is
anisotropic XY model, which encompasses two well-known spin models,
i.e., Ising chain and the XX (isotropic XY) chain in a transverse
field. The XY model holds an advantage that it can be exactly solved
by mapping to a spinless fermionic model. Such solvability provides
a playground for testing the physical ideas \cite{wenlong}. For
one-qubit case, Quan \textit{et al.} \cite{quan} first proved that
the decay of the Loschmidt echo (LE) was enhanced by the QPT of the
Ising bath. With that, this consideration was extended to the
two-qubit case. In the transverse Ising model, Sun \textit{et al.
}\cite{sun07} showed that the concurrence decayed exponentially with
fourth power of time in the vicinity of the critical point of spin
bath. For an XY spin chain with Dzyaloshinsky-Moriya (DM)
interaction, Cheng \textit{et al.} \cite{cheng} found that decay of
decoherence factor was sensitive to the DM interaction, especially
in the strong-coupling region. Also, the three-qubit case was
studied by some works \cite{ma,ma08}. The previous works have
extensively studied the decoherence process of initially entangled
state of noninteracting qubits \cite{sun07,cheng,ma,ma08}.

In this paper, we investigate the influence of the XY spin bath with
DM interaction term for two interacting qubits. The paper is
organized as follows. In Sec. \ref{MODEL}, we introduce the model
Hamiltonian describing two-qubit system coupled to an XY spin chain
with DM interaction, and derive
an analytic formula for the exact solution of entanglement. In Sec. \ref%
{Entanglementdynamics}, regarding the spin bath as a decoherent environment,
we exhibit the competition between system-bath and interqubit interaction in
the entanglement dynamics. We analyze the competing effects of environmental
noise and two-qubit interaction for various system parameters, especially at
the QCP of spin bath. We observe resonance peaks emerging for some specific
parameters. Finally, we give a summary of our results in Sec. \ref%
{Conclusions}.
\begin{figure}[h]
\includegraphics[width=8cm]{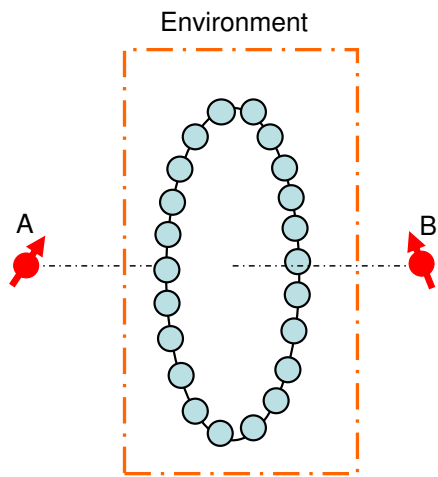}
\caption{(Color online) The schematic diagram of two external
interacting spin qubits symmetrically coupled to the environment
described as an XY spin chain with Dzyaloshinsky-Moriya term.}
\label{fig:schematic}
\end{figure}
\section{Model and solution}

\label{MODEL} The quantum system we consider consists of two
interacting qubits $s_{A}^{z}$ and $s_{B}^{z}$ coupled to a spin
environment $E$. The environment is composed by $N$ spin-1/2
particles with periodic boundary condition, which can be described
by one-dimensional XY spin chain with DM interaction
\cite{Lieb-AnnPhys-1961}. The schematic diagram of the quantum open
system is shown in Fig. \ref{fig:schematic}. The corresponding
Hamiltonian is given by
\begin{eqnarray}
H=H_{E}^{\lambda }+H_{AB}+H_{I},  \label{MainHamiltonian}
\end{eqnarray}
with
\begin{eqnarray}
H_{E}^{\lambda }&=&\sum_{j=1}^{N}\left( \frac{1+\gamma }{2}\sigma
_{j}^{x}\sigma _{j+1}^{x}+\frac{1-\gamma }{2}\sigma _{j}^{y}\sigma
_{j+1}^{y}+\lambda \sigma _{j}^{z}\right)  \notag \\
&&+\sum_{j=1}^{N}\vec{D}\cdot (\vec{\sigma}_{j}\times \vec{\sigma}_{j+1}), \\
H_{AB}&=&Js_{A}^{z}s_{B}^{z}, \\
H_{I}&=&g(s_{A}^{z}+s_{B}^{z})\sum_{j=1}^{N}\sigma _{j}^{z},
\end{eqnarray}
where $\sigma _{j}^{a}$ ($a=x$, $y$, $z$) is the Pauli matrix at
site $j$. The parameters $\gamma $ and $\lambda $ characterize the
anisotropy of the
self-Hamiltonian and the intensity of the magnetic field applied along the $%
z $ axis, respectively. Here, such self-Hamiltonian $H_{E}^{\lambda
}$ comprises DM term, which is an antisymmetric spin coupling. The
DM interaction often arises from a mixture of superexchange and
spin-orbit coupling in low-dimensional magnetic materials
\cite{Dzyaloshinsky,Moriya}. For analytical solvability, we assume
that the $\vec{D}$ vector is imposed along the $z$ direction., i.e.,
$\vec{D}=D\vec{z}$. The strength of interaction between the system
qubits is given by $J(>0)$, and the coupling between the system
qubits and the surrounding spin chain is denoted by $g$.

Since $[H_{AB},H_{I}]=0$, Hamiltonian (\ref{MainHamiltonian}) can be
rewritten as
\begin{equation}
H=\sum_{\mu=1}^{4}\varepsilon_{\mu}|\phi_{\mu}\rangle\langle\phi_{\mu}|
\otimes H_{E}^{\lambda_{\mu}},
\end{equation}
where $|\phi_{\mu}\rangle$ $(\mu=1,\ldots,4)$ is the $\mu$th eigenstate of $%
H_{AB}$ with eigenvalue $\varepsilon_{\mu}$. Under these bases, the
coupling
$H_{I}$ between the qubits and spin chain exerts an extra magnetic field $%
\xi_{\mu}$ on the spin bath $H_{E}^{\lambda}$, and then gives rise
to an effective qubit-dressed Hamiltonian $H_{E}^{\lambda_{\mu}}$
with magnetic field $\lambda_{\mu}=\lambda+\xi_{\mu}$. It is easy to
find that
\begin{equation}
\xi_{1}=g,\quad\xi_{2}=\xi_{3}=0,\quad\xi_{4}=-g.  \label{xi}
\end{equation}

The Hamiltonian $H_{E}^{\lambda _{\mu }}$ can be diagonalized by following
the standard procedures. The Jordan-Wigner transformation \cite%
{Jordan-Zphys-1928,Sachdev}
\begin{align}
\sigma _{j}^{+}& =\exp \left[ i\pi \sum_{i=1}^{j-1}f_{i}^{\dagger }f_{i}%
\right] f_{j}=\prod_{i=1}^{j-1}\sigma _{i}^{z}f_{j},  \notag \\
\sigma _{j}^{-}& =\exp \left[ -i\pi \sum_{i=1}^{j-1}f_{i}^{\dagger }f_{i}%
\right] f_{j}^{\dagger }=\prod_{i=1}^{j-1}\sigma
_{i}^{z}f_{j}^{\dagger },
\notag \\
\sigma _{j}^{z}& =1-2f_{j}^{\dagger }f_{j},
\end{align}%
maps spin chain to one-dimensional spinless fermionic model with creation
and annihilation operators as follows
\begin{align}
H_{E}^{\lambda _{\mu }}&
=\sum_{j=1}^{N-1}\Bigg[(1-2iD)f_{j}^{\dagger
}f_{j+1}+(1+2iD)f_{j+1}^{\dagger }f_{j}  \notag \\
& +\gamma \left( f_{j}^{\dagger }f_{j+1}^{\dagger
}+f_{j+1}f_{j}\right)
+\lambda _{\mu }\left( 1-2f_{j}^{\dagger }f_{j}\right) \Bigg]  \notag \\
& +\Bigg[(1-2iD)f_{N}^{\dagger }f_{1}+(1+2iD)f_{1}^{\dagger }f_{N}  \notag \\
& +\gamma \left( f_{N}^{\dagger }f_{1}^{\dagger }+f_{1}f_{N}\right) \Bigg]%
 \exp (i\phi )+\lambda _{\mu }\left( 1-2f_{N}^{\dagger }f_{N}\right) .
\end{align}%
An extra phase $\phi$ = $\pi(\sum_{j=1}^{N}f_{j}^{\dagger }f_{j}+1)
$ appears on the chain boundary  due to phase accumulation of the
Jordan-Wigner transformation.

Next discrete Fourier transformation is introduced to convert the fermionic
operators from real space to momentum space by defining
\begin{equation}
d_{k}=\frac{1}{\sqrt{N}}\sum_{j}e^{ikj}f_{j},\text{ \ \ } d_{k}^{\dagger }=%
\frac{1}{\sqrt{N}}\sum_{j}e^{-ikj}f_{j}^{\dagger },
\end{equation}
with the discrete momentums as
\begin{equation}
k=\frac{(2n\pi+\phi)}{N}, \quad n=-\frac{N}{2}, -\frac{N}{2}+1, ...,
\frac{N}{2}-1.
\end{equation}%
After that, the Hamiltonian becomes
\begin{eqnarray}
H_{E}^{\lambda _{\mu }} &=&\sum_{k}\left( 2\cos k-2\lambda _{\mu }-4D\sin
k\right) d_{k}^{\dagger }d_{k}  \notag \\
&&+i\gamma \sin k\left( d_{-k}^{\dagger }d_{k}^{\dagger
}+d_{-k}d_{k}\right) +N\lambda _{\mu }.
\end{eqnarray}%
The diagonalized form is achieved by Bogoliubov transformation which defines
quasiparticle creation (annihilation) operator $b_{k,\lambda _{\mu
}}^{\dagger }$ ($b_{k,\lambda _{\mu }}$) as
\begin{align}
d_{k}& =\cos \frac{\theta _{k}^{\lambda _{\mu }}}{2}b_{k,\lambda
_{\mu }}+i\sin \frac{\theta _{k}^{\lambda _{\mu }}}{2}b_{-k,\lambda
_{\mu
}}^{\dagger },  \notag \\
d_{k}^{\dagger }& =\cos \frac{\theta _{k}^{\lambda _{\mu
}}}{2}b_{k,\lambda
_{\mu }}^{\dagger }-i\sin \frac{\theta _{k}^{\lambda _{\mu }}}{2}%
b_{-k,\lambda _{\mu }},
\end{align}%
with the angle $\theta _{k}^{\lambda _{\mu }}$ defined by
\begin{equation}
\tan \theta _{k}^{\lambda _{\mu }}=\frac{-\gamma \sin k}{\lambda
_{\mu }-\cos k}. \label{angle}
\end{equation}%
Consequently, the dressed self-Hamiltonian is unitarily equivalent to such
diagonal form
\begin{equation}
H_{E}^{\lambda _{\mu }}=\sum_{k}\Omega _{k}^{\lambda _{\mu }}(b_{k,\lambda
_{\mu }}^{\dagger }b_{k,\lambda _{\mu }}-\frac{1}{2}),
\end{equation}%
where
\begin{equation}
\Omega _{k}^{\lambda _{\mu }}(k)=2\sqrt{\left( \lambda _{\mu }-\cos k\right)
^{2}+\left( \gamma \sin k\right) ^{2}}-4D\sin k.
\end{equation}%
The ground state $|G\rangle _{\lambda _{\mu }}$ has no quasiparticle
for arbitrary $k$, i.e., $b_{k,\lambda _{\mu }}|G\rangle _{\lambda
_{\mu }}=0$. Due to the relation $b_{k,\lambda _{\mu }}=\cos(\theta
_{k}^{\lambda _{\mu
}}/2) $ $d_{k}-$$i\sin (\theta _{k}^{\lambda _{\mu }}/2)d_{-k}^{\dagger }$, $%
|G\rangle _{\lambda _{\mu }}$ can be written as
\begin{equation}
|G\rangle _{\lambda _{\mu }}=\prod_{k>0}\left( \cos \frac{\theta
_{k}^{\lambda _{\mu }}}{2}|0\rangle _{k}|0\rangle _{-k}+i\sin \frac{\theta
_{k}^{\lambda _{\mu }}}{2}|1\rangle _{k}|1\rangle _{-k}\right) ,
\end{equation}%
where $|0\rangle _{k}$ and $|1\rangle _{k}$ are the vacuum and
single excitation of the $k$th mode $d_{k}$, respectively. With
these analytical expressions, we can straightforwardly derive the
time evolution of arbitrary initial state and obtain the reduced
density matrix of the two-qubit system, and then examine the effect
of the environment. Suppose that at time $t=0$ the qubits are
completely disentangled from the environment, i.e., the global
system wave function is given by
\begin{equation}
|\Psi (0)\rangle =|\phi (0)\rangle _{AB}\otimes |\phi (0)\rangle _{E}.
\end{equation}

Clearly, governed by the Hamiltonian (\ref{MainHamiltonian}), the state at
time $t$ is given by $|\Psi(t)\rangle=U(t)|\Psi(0)\rangle$, where $%
U(t)=\exp(-iHt)$ is the evolution operator of the composite system. The
reduced density matrix of the two-qubit system is obtained
\begin{align}
\rho_{A,B} & =\text{\textrm{Tr}}_{E}(|\Psi(t)\rangle\langle\Psi (t)|)  \notag
\\
& =\text{\textrm{Tr}}_{E}(e^{-iHt}|\Psi(0)\rangle\langle\Psi(0)|e^{iHt})
\notag \\
& =\sum_{\mu,\nu=1}^{4}c_{\mu}(t)c_{\nu}^{\ast}(t)F_{\mu\nu}(t)|\phi_{\mu
}\rangle\langle\phi_{\nu}|,  \label{eqn:Fmunv}
\end{align}
where

\begin{equation}
c_{\mu}=e^{-i\varepsilon_{\mu}t}
\langle\phi_{\mu}|\Phi(0)\rangle_{AB},
\end{equation}
and the decoherence factors are

\begin{equation}
F_{\mu\nu}(t)={}_{E}\langle\phi(0)|\exp(iH_{E}^{\lambda_{\nu}}t)\exp
(-iH_{E}^{\lambda_{\mu}}t)|\phi(0)\rangle_{E}.
\end{equation}

We assume that the two qubits in $AB$ system initially stem from a separable
state, i.e., $|\phi(0)\rangle_{AB}=(|0\rangle+|1\rangle)/\sqrt{2}%
\otimes(|0\rangle+|1\rangle)/\sqrt{2}$, where $|0\rangle$ and
$|1\rangle$ denote the spin up and down, respectively. The initial
state of the
environment is supposed as the ground state of $H_{E}^{\lambda}$, i.e., $%
|\phi(0)\rangle_{E}=|G\rangle_{\lambda}$. By tedious calculation
\cite{cheng,yuan07}, we have
\begin{align}
F_{\mu\nu}(t) & ={}_{\lambda}\langle
G|e^{iH_{E}^{\lambda_{\nu}}t}e^{-iH_{E}^{\lambda_{\mu}}t}|G\rangle_{\lambda}
\notag \\
&
=\prod_{k>0}\cos\Theta_{k}^{\lambda_{\mu}}\cos\Theta_{k}^{\lambda_{%
\nu}}e^{i(\Omega_{k}^{\lambda_{\mu}}-\Omega_{k}^{\lambda_{\nu}})t}\cos(%
\Theta _{k}^{\lambda_{\mu}}-\Theta_{k}^{\lambda_{\nu}})  \notag \\
&
+\sin\Theta_{k}^{\lambda_{\mu}}\sin\Theta_{k}^{\lambda_{\nu}}e^{-i(%
\Omega_{k}^{\lambda_{\mu}}-\Omega_{k}^{\lambda_{\nu}})t}\cos(\Theta
_{k}^{\lambda_{\mu}}-\Theta_{k}^{\lambda_{\nu}})  \notag \\
&
+\sin\Theta_{k}^{\lambda_{\mu}}\cos\Theta_{k}^{\lambda_{\nu}}e^{-i(%
\Omega_{k}^{\lambda_{\mu}}+\Omega_{k}^{\lambda_{\nu}})t}\sin(\Theta
_{k}^{\lambda_{\mu}}-\Theta_{k}^{\lambda_{\nu}})  \notag \\
&
-\cos\Theta_{k}^{\lambda_{\mu}}\sin\Theta_{k}^{\lambda_{\nu}}e^{i(%
\Omega_{k}^{\lambda_{\mu}}+\Omega_{k}^{\lambda_{\nu}})t}\sin(\Theta
_{k}^{\lambda_{\mu}}-\Theta_{k}^{\lambda_{\nu}}),
\label{coherencefactor1}
\end{align}
where $\Theta_{k}^{\lambda_{\mu}}=(\theta_{k}^{\lambda_{\mu}}-\theta
_{k}^{\lambda})/2$ is the angle difference between the normal mode
dressed by the system-environment interaction and the purely
environment.

From Eq. (\ref{eqn:Fmunv}), in the bases spanned by
$\{|00\rangle,|01\rangle,|10\rangle,|11\rangle\}$, the reduced
density matrix of the two-qubit system is obtained

\begin{equation}
\rho _{AB}(t)=\frac{1}{4}{\small \left(
\begin{array}{cccc}
1 & e^{-\frac{iJt}{2}}F_{12}(t) & e^{-\frac{iJt}{2}}F_{13}(t) & F_{14}(t) \\
e^{\frac{iJt}{2}}F_{12}^{\ast }(t) & 1 & F_{23}(t) & e^{\frac{iJt}{2}%
}F_{24}(t) \\
e^{\frac{iJt}{2}}F_{13}^{\ast }(t) & F_{23}^{\ast }(t) & 1 & e^{\frac{iJt}{2}%
}F_{34}(t) \\
F_{14}^{\ast }(t) & e^{-\frac{iJt}{2}}F_{24}^{\ast }(t) & e^{-\frac{iJt}{2}%
}F_{34}^{\ast }(t) & 1%
\end{array}%
\right).}  \label{rho}
\end{equation}

Some matrix elements can be further simplified for the choice of $%
\lambda _{\mu }$. For the case of Eq. (\ref{xi}), we have $\Theta
_{k}^{\lambda _{2}}=\Theta _{k}^{\lambda _{3}}=0$, and also it is obvious
that such relations hold: $F_{12}=F_{13}$, $F_{23}=1$, $F_{24}=F_{34}$. So
there are three independent decoherence factors.

To investigate the entanglement dynamics of two qubits surrounding spin
bath, rather than decoherence factor employed in Refs. \cite{cheng,yuan07},
we utilize the concurrence directly \cite{Wootters}, an entanglement measure
for any bipartite system that relates to the two-site reduced density matrix
$\rho$. The concurrence for two qubits is defined as
\begin{equation}
C(t)=\max\{0,\sqrt{\omega_{1}}-\sqrt{\omega_{2}}-\sqrt{\omega_{3}}-\sqrt{%
\omega_{4}}\},  \label{ConcurrenceDefinition}
\end{equation}
where $\omega_{i}$ $(i=1,\ldots,4)$ are the eigenvalues in
decreasing order of the auxiliary
matrix $\zeta=\rho(\sigma_{y}\otimes\sigma_{y})\rho^{\ast}(\sigma_{y}\otimes%
\sigma_{y})$. Here $\rho^{\ast}$ denotes the complex conjugation of $\rho$
in the standard bases and $\sigma_{y}$ is the Pauli matrix. The concurrence
varies from $C=0$ for a separable state (zero entanglement) to $C=1$ for a
maximally entangled state.

\section{Entanglement dynamics of the two-qubit system}

\label{Entanglementdynamics}

\begin{figure}[t]
\includegraphics[width=0.5\textwidth]{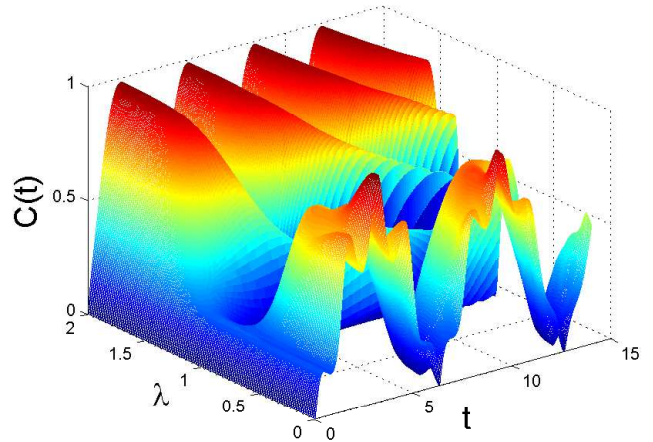}
\caption{(Color online) Concurrence evolution versus $\protect\lambda $ for $%
N=801$, $J=2$, $\protect\gamma =1$, $D=0$ when there is
weak-coupling between qubits and spin bath, i.e., $g=0.05$.}
\label{weakcoupling-sys-env}
\end{figure}

\begin{figure}[tbp]
\includegraphics[width=0.5\textwidth]{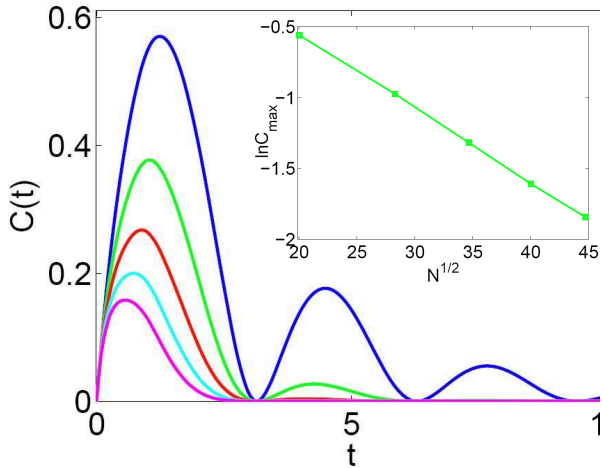}
\caption{(Color online) Concurrence versus time $t$ is plotted for
different chain length $N$ with $\lambda=1$, g=0.05, $\gamma$=1,
J=2, D=0. From above to below, the lines correspond to $N$= 401,
801, 1201, 1601 and 2001. The logarithm of maximum concurrence
versus $\sqrt{N}$ is shown in the inset.} \label{fig:C-N}
\end{figure}

Before we consider the competition of system-bath coupling and two-qubit
interaction, we first emphasize the following two limiting cases. On one
hand, when there is no coupling between system and environment, i.e., $g=0$,
we can get $\omega_{1}=\sin^{2}Jt/2$, $\omega_{2}=\omega_{3}=%
\omega_{4}=0 $, and then $C(t)=|\sin Jt/2|$. It means the
interaction
between the qubits can generate an oscillating entanglement with period of $%
2\pi/J$. At $t=(2m+1)\pi/J$ ($m=0,1,2,...$), the state of the system
reaches the maximum entanglement. On the other hand, when there is
no interaction between the two qubits, the entanglement dynamics of
the qubit system returns to the process of entanglement induced by
the spin bath \cite{yi}. The entanglement changes dramatically along
the line of critical points of the spin bath. Now we investigate the
influence of the spin bath during the process of entanglement
generation. In Fig. \ref{weakcoupling-sys-env}, concurrence is
plotted as a function of magnetic field $\lambda$ and time $t$ with
$J=2,\gamma=1,D=0$ in weak-coupling ($g\ll1$) region. It shows that
the spin bath will weaken the entanglement generation when the
coupling between system and environment is adiabatically turned on.
Especially, the quantum phase transition of XY model at $\lambda=1$
will greatly enhance the decoherence, which coincides with the rapid
decay of LE in an Ising model \cite{quan}. It should be noted that
we consider dynamics evolution of the system based on a finite-sized
environment. To examine the effect of the chain length $N$ on the
quantum entanglement evolution, we plot Fig. \ref{fig:C-N}. It shows
that the longer the spin chain, the more serious the attenuation at
QCP. The concurrence displays oscillatory decay of time for $N=401$
and $801$. As the length of the chain increases, the maximum value
of concurrence decreases, and the revival of the concurrence
disappears. The inset of Fig. \ref{fig:C-N} shows that the maximum
concurrence decays exponentially with the square root of chain
length $N$. In addition, in weak-coupling region, for large
$\lambda$, the concurrence restores the sine function versus time.
It seems that the effect of system-bath coupling is insignificant.
The reason is that the spin bath is polarized along $z$ axis in
strong magnetic field. In this case, each spin is not entangled with
the rest spins and the qubits \cite{sjgu1,sjgu2}. In other word, the
ferromagnetic bath can be more or less thought of as classical.
Therefore, it has negligible effect on qubits, and the interaction
between qubits plays a dominant role in entanglement generation
again, as shown in Fig. \ref{weakcoupling-sys-env}.

\begin{figure}[t]
\includegraphics[bb=25 241 581 630,width=0.5\textwidth]{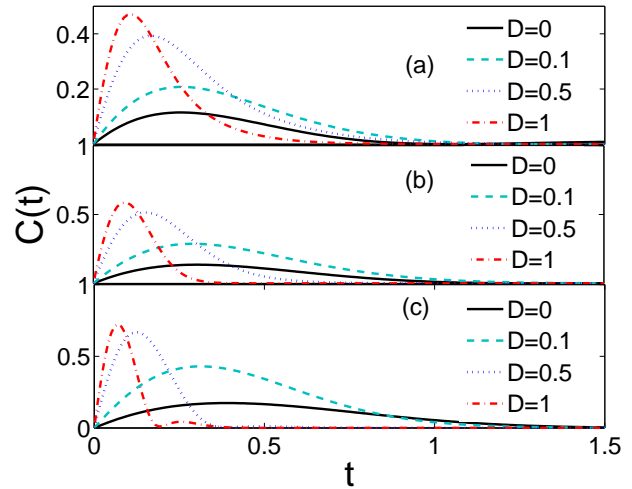}
\caption{(Color online) Concurrence as a function of time $t$ under
different $D$ with (a) $\protect\gamma =0.8$, (b) $\protect\gamma
=0.6$, and
(c) $\protect\gamma =0.4$. We set other parameters as $\protect\lambda =1.0$%
, $J=2$, $g=0.05$, and $N=2001$.} \label{DM}
\end{figure}

Then we investigate the effect of DM interaction at QCP of the spin bath. In
Fig. \ref{DM} (a)-(c), we display the concurrence against time $t$ under
different $D$ with $\gamma$ equals to $0.8$, $0.6$ and $0.4$, respectively.
We can see the maximum concurrence is enhanced by increasing $D$. The bigger
the magnitude of $D$, the less time needed to reach the maximum concurrence.
The growth of entanglement can be interpreted via the fact that the DM
interaction arouses the strong planar quantum fluctuations. We would like to
point out that the entanglement generation is suppressed at the QCP of spin
bath as in Fig. \ref{weakcoupling-sys-env}, so the maximum concurrence can
not reach unity. However, as the anisotropy parameter $\gamma$ decreases,
the quantum fluctuations arising from the in-planar interactions become
stronger \cite{kar}, and then the two qubits in the presence of quantum
fluctuations are more quantum correlated. In a word, the suppression of the
entanglement at the QCP can be compensated both by increasing the DM
interaction and by decreasing the anisotropy of the spin chain.

\begin{figure}[t]
\includegraphics[width=0.5\textwidth]{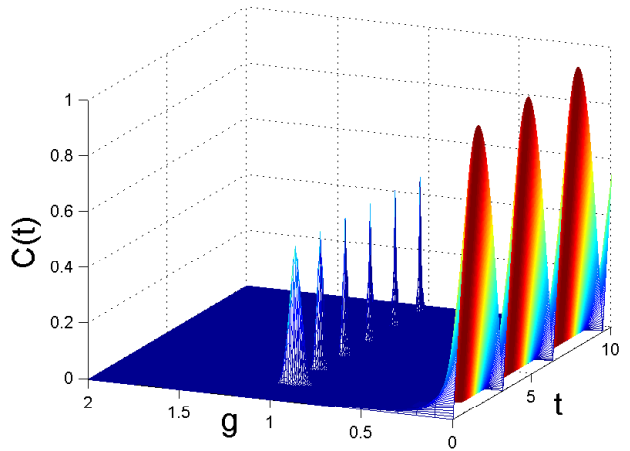}
\caption{(Color online) Concurrence as a function of time $t$ and
coupling
parameter $g$ with $N=801$, where we set other parameters as $\protect%
\lambda =1.0$, $J=2.0$, $\protect\gamma =1$, and $D=0$.} \label{g}
\end{figure}

\begin{figure}[t]
\includegraphics[bb=-30 220 608 654,width=0.45\textwidth]{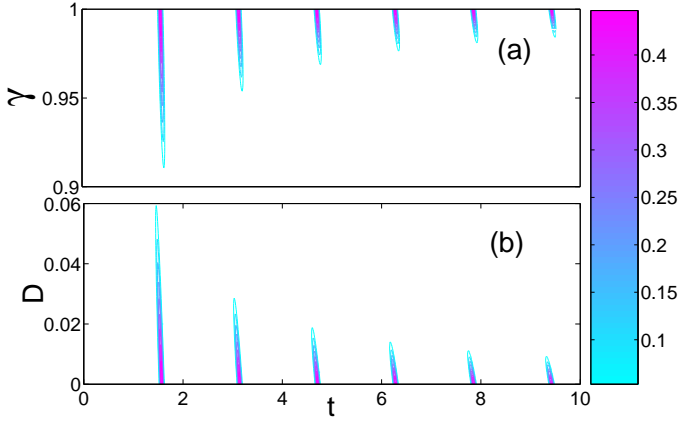}
\caption{(Color online) Concurrence as a function of time $t$ and
(a) the anisotropy parameter $\protect\gamma $, (b) the intensity of
the DM
interaction $D$. We set other parameters as $N=801$, $g=\protect\lambda =1.0$%
, $J=2.0$ with (a) $D=0$ and (b) $\protect\gamma =1$.}
\label{Dandgamma}
\end{figure}

We now turn to the entanglement dynamics beyond the weak-coupling region.
Fig. \ref{g} shows the concurrence as a function of time and the coupling $g$%
. When the coupling $g$ increases, due to the decoherence of spin
environment, the concurrence damps dramatically. Thus in the strong-coupling
region, no entanglement is generated. However, it should be noted that in
the vicinity of $g=\lambda=1$, the suppression of the entanglement
generation at the quantum critical point is released. At some time
entanglement emerges suddenly and reaches a maximum value in a short time.
It can be easily concluded from Fig. \ref{Dandgamma} that the resonance
peaks periodically appear at $t=m\pi/2$, where $m$ is an integer. We study
the behavior of the resonance peaks with respect to parameters $D$ and $%
\gamma$. As shown in Fig. \ref{Dandgamma} (a), resonance peaks
disappear quickly as $\gamma$ decreases, and there is no revival
when $\gamma$ is less than certain threshold. A similar phenomenon
builds up as the parameter $D$ increases, as shown in Fig.
\ref{Dandgamma} (b). This reveals that these resonance peaks are
highly sensitive to both the anisotropy parameter and the DM
interaction strength.

\begin{figure}[t]
\includegraphics[bb=0 350 680 630,width=0.5\textwidth]{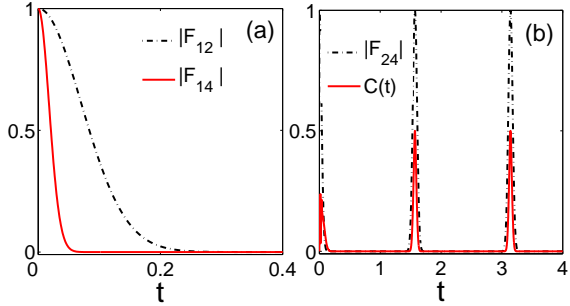}
\caption{(Color online) Docoherence factor $|F_{\protect\mu
\protect\nu }|$
and concurrence as a function of time $t$. We set other parameters as $N=801$%
, $g=\protect\lambda =1.0$, $J=2.0$, $\ D=0$ and $\protect\gamma
=1$.} \label{g=1}
\end{figure}

In order to make a scrutiny into the resonance peaks, we rule out the effect
of anisotropy and DM terms, i.e., $D=0$, $\gamma =1$. Now we have $\lambda
_{1}=2\lambda $, $\lambda _{2}$ =$\lambda _{3}$ = $\lambda $, and $\lambda
_{4}=0$. Under these parameters, Fig. \ref{g=1} depicts the decoherence
factors and concurrence against time in the resonance case $g=\lambda =1$.
We can see that the decoherence factors $\left\vert F_{12}\right\vert $ and $%
\left\vert F_{14}\right\vert $ collapse to zero fleetly, while $\left\vert
F_{24}\right\vert $ displays periodic revivals as time goes on. According to
Eq. (\ref{ConcurrenceDefinition}), the concurrence is dominated by $%
\left\vert F_{24}\right\vert $ when $\left\vert F_{12}\right\vert $ and $%
\left\vert F_{14}\right\vert $ decay to zero, and exhibits the
similar behavior as $\left\vert F_{24}\right\vert $.

To illustrate why there are robust peaks of concurrence for $\gamma=1$ and $%
D=0$, let us give a heuristical explanation as follows. By Eq. (\ref%
{coherencefactor1}), we have
\begin{eqnarray}
F_{12}(t)=\prod_{k>0}e^{-i\Omega_{k}^{\lambda_{2}}t}\left[ \cos
^{2}(\Theta_{k}^{\lambda_{1}})e^{i\Omega_{k}^{\lambda_{1}}t}+\sin^{2}(%
\Theta_{k}^{\lambda_{1}})e^{-i\Omega_{k}^{\lambda_{1}}t}\right].\nonumber
\\
\end{eqnarray}
Such form of expression has been studied carefully in Refs. \cite%
{cheng,Cucchietti1,Cucchietti2}. $|F_{12}(t)|$ shows a Gaussian decay with
time for $\lambda = 1$. Interestingly, the decoherence factor $%
F_{14}(t)=$ \\ ${}_{\lambda}\langle
G|e^{iH_{E}^{\lambda_{4}}t}e^{-iH_{E}^{\lambda_{1}}t}|G\rangle_{\lambda}$
is
similar to the decoherence factor in the spin-echo experiment in Ref. \cite%
{Cucchietti2}, and also the decay of $|F_{14}(t)|$ is Gaussian
\cite{cheng}, as shown in Fig. \ref{g=1}(a).

A remarkable difference arises from $F_{24}(t)$. From Eq. (\ref%
{coherencefactor1}), we have
\begin{align}
F_{24}(t) & =\prod_{k>0}e^{i\Omega_{k}^{\lambda_{2}}t}\left[ \cos
^{2}(\Theta_{k}^{\lambda_{4}})e^{-i\Omega_{k}^{\lambda_{4}}t}+\sin^{2}(%
\Theta_{k}^{\lambda_{4}})e^{i\Omega_{k}^{\lambda_{4}}t}\right] .
\notag \\ & \label{Eq:F24}
\end{align}
Due to the condition that $g=\lambda$, such Hamiltonian has an important
feature: the spectra $\Omega_{k}^{\lambda_{4}}$ of all the modes are
independent of momentum, i.e., $\Omega_{k}^{\lambda_{4}}=2$. Consequently,
at $t=m\pi/2$ ($m=$0, 1, 2, $\ldots$), $F_{24}$$(m\pi/2)$=$\prod_{k>0}$$(-1)^{m}$$e^{im%
\Omega _{k}^{\lambda_{2}}\pi/2}$. The norm of $F_{24}(m\pi/2)$ is
unity. Based on such analysis, at $t =m\pi/2$, we can approximatively set $%
F_{12}=0$, $F_{14}=0$, $F_{24}=1$ in Eq. (\ref{rho}), and thus we obtain $%
C(m\pi/2)=0.5$. For a tiny time deviation $\delta$ from resonance
time $m\pi/2$, it arrives
\begin{align}
|F_{24}(\delta)| & =\prod_{k>0}\left\vert
\cos^{2}(\Theta_{k}^{\lambda_{4}})e^{-2i\delta}+\sin^{2}(\Theta_{k}^{%
\lambda_{4}})e^{2i\delta}\right\vert  \notag \\
& \simeq e^{-A\delta^{2}},
\end{align}
with $A=\sum_{k>0}(1-\cos(4\Theta_{k}^{\lambda_{4}}))$. In other
words, the decoherence factor $F_{24}(t)$ will exponentially decay
in a short time after deviating from the resonance time. Therefore,
at $t \neq m\pi/2$, replacing $F_{12}$, $F_{14}$ and $F_{24}$ with
zeros in Eq. (\ref{rho}), we get $C(t)=0$. It means that the
entanglement generation is due to the presence of the
momentum-independent spectrum structure. The flat band will also be
destroyed when $\gamma$ is deviated from $1$ or $D$ is turned on, so
the resonance peaks will be suppressed by varying $\gamma$ or $D$,
as shown in Fig. \ref{Dandgamma}. These interesting features are
also confirmed in our numerical study. With regard to the parameters
of $\gamma=1$, $D=0$ and $g=\lambda$, we obtain
\begin{eqnarray}
 A=\sum_{k>0}\frac{2\lambda^2 \sin^2 k}{\lambda^2 -
2\lambda \cos k +1}  = \left\{ \begin{array}{ll }
\frac{N \lambda^2}{2},  \quad & \lambda < 1,\\
\frac{N}{2}, \quad & \lambda \ge 1.
\end{array} \right.
\end{eqnarray}
Thus, the width of spikes of $\vert F_{24}\vert$ in Fig.
\ref{g=1}(b) is inversely proportional to $\sqrt{N}\lambda$ when
$\lambda <1$, while it is inversely proportional to $\sqrt{N}$ when
$\lambda \ge 1$. Since the concurrence has the same dependence as
$F_{24}$, the resonance peaks at QCP will get sharper with the
increase of chain length $N$. The peaks are robust even at the
thermodynamic limit.

\begin{figure}[t]
\includegraphics[bb=-80 196 700 620,width=0.48\textwidth]{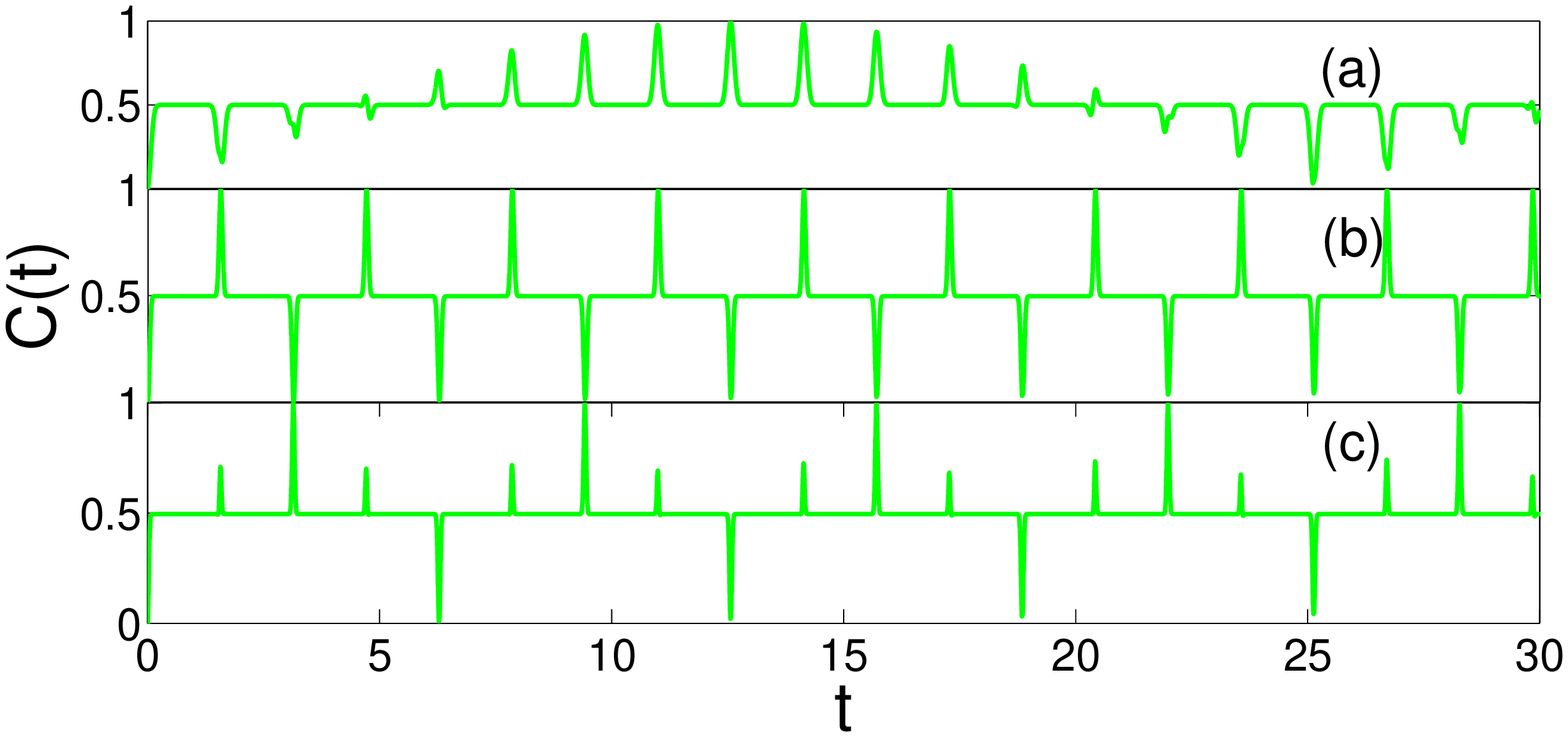}
\includegraphics[bb=-80 260 700 600,width=0.48\textwidth]{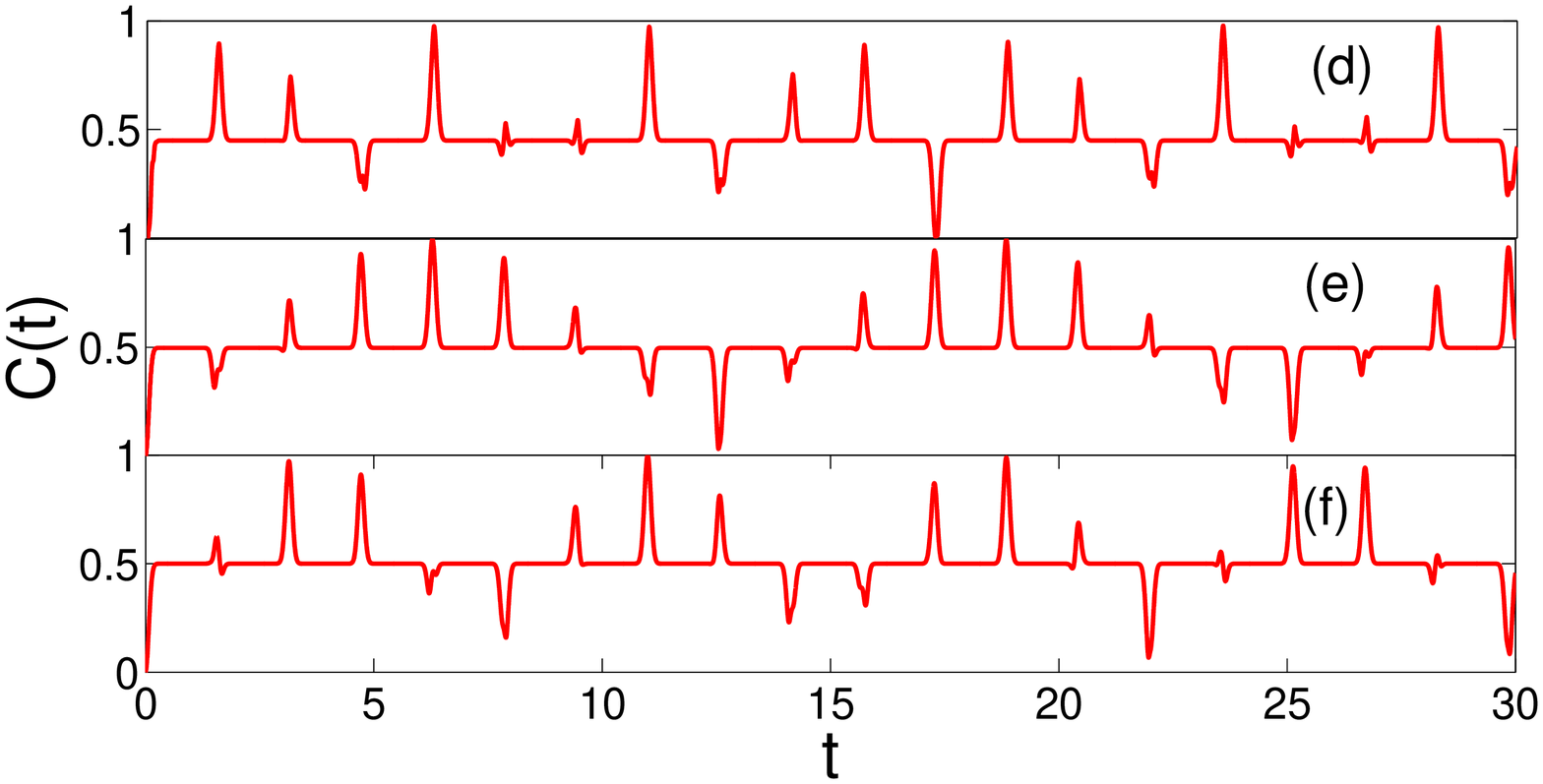}
\caption{(Color online) Concurrence as a function of time $t$ for
(a) N=201, (b) N=801, (c) N=1601 with $g$=$\lambda$=100, $J=2$,
$\protect\gamma =1$, and $D=0$; (d) $g$=$\lambda$=10, (e)
g=$\lambda$=50, (f) g=$\lambda$=500 with $N=201$, $J=2$,
$\protect\gamma =1$, and $D=0$.} \label{g=lambda}
\end{figure}

We want to emphasize that the behavior of $\vert F_{24}(t) \vert$ is
universal for any $g=\lambda$, while the Gaussian decay for
$F_{12(14)}(t)$ is only shown around QCP. When $g=\lambda \gg 1$,
from the angle for Bogoliubov transformation in Eq. (\ref{angle}),
we have
\begin{eqnarray}
\theta_{k}^{\lambda}&\approx&0, \quad \theta_{k}^{\lambda_{1}}
\approx0, \quad \theta_{k}^{\lambda_{2}}\approx 0,
\quad \theta_{k}^{\lambda_{3}}\approx0,  \\
\theta_{k}^{\lambda_{4}}&\approx& \left\{
\begin{array}{ll }
\pi + k, \quad & -\pi < k < 0,\\
\pi - k, \quad &  0 \le k  \le \pi.
\end{array} \right.
\end{eqnarray}
Consequently,
\begin{eqnarray}
F_{12}(t)& \approx &\prod_{k>0}e^{i(\Omega_{k}^{\lambda_{1}}
-\Omega_{k}^{\lambda_{2}})t}, \\
F_{14}(t)& \approx &\prod_{k>0}e^{i\Omega_{k}^{\lambda_{1}}t}\left[
\cos
^{2}(\Theta_{k}^{\lambda_{4}})e^{-i\Omega_{k}^{\lambda_{4}}t}+\sin^{2}(%
\Theta_{k}^{\lambda_{4}})e^{i\Omega_{k}^{\lambda_{4}}t}\right].
\nonumber \\ \label{Eq:F14}
\end{eqnarray}
Apparently, $\vert F_{12}(t)\vert \approx 1$, $F_{14}(t)/F_{24}(t)
\approx F_{12}(t)$. It is notable that the form of Eq.
(\ref{Eq:F14}) is similar with Eq. (\ref{Eq:F24}). Compared with
$\vert F_{24} (t) \vert $ analyzed above, $\vert F_{12} (t) \vert$
should behave analogically, and spikes also spring up at $t
=m\pi/2$. The situation is contrary to the case of $g=\lambda = 1$.
At $t \neq m\pi/2$, all the decoherence factors are zeros except
$F_{12}(t)$, so the concurrence is dominated by $\vert
F_{12}(t)\vert$, and thus the concurrence forms a flat platform
structure, which remains at $C(t)=0.5$. As for $t= m\pi/2$, the
concurrence shows exotic patterns, which are sensitive to the chain
length $N$ and the amplitude of coupling $g$. Peaks and valleys emit
from the platform at $t= m\pi/2$, where the system can reach maximum
or zero concurrence state periodically, as depicted in
Fig.\ref{g=lambda}. According to the analysis above, we know that
the width of peaks and valleys is inversely proportional to
$\sqrt{N}$. It is observed that the width of peaks and valleys
decreases as the chain length $N$ increase (see Fig. \ref{g=lambda}
(a)-(c)), and it is irrelevant to the coupling strength $g$ (see
Fig. \ref{g=lambda} (d)-(f)). For the region out of scope of QCP and
strong coupling with $g=\lambda$, there exists competitive effect
among these three decoherence factors.
\section{Conclusions}

\label{Conclusions} The investigation of decoherence of spin environment is
an attractive topic in the entanglement dynamical process. In the paper, we
have studied the entanglement dynamics in term of concurrence of two
interacting spin qubits coupled to a general XY spin chain with DM
interaction. When there is no coupling between two-qubit system and
environment, the interaction between the qubits will generate entanglement
as a sine function versus time. In the weak-coupling region, the
entanglement generation is suppressed remarkably near the QCP of the spin
bath. However, this suppression can be compensated both by increasing the DM
interaction or by decreasing the anisotropy of the spin chain. Beyond the
weak-coupling region, the system-bath coupling will make the entanglement in
the system decay rapidly except for $g=\lambda $, where periodic resonance
peaks of concurrence appear. By analyzing the decoherence factors, we
attribute the appearance of resonance peaks to the emergence of the constant
energy flat band. In such circumstances, increasing DM interaction or
decreasing anisotropy will spoil the flatness of the band, and then the
resonance peaks collapse. The effects of $\gamma $ and $D$ here seem
contrary to case of weak coupling. The seemingly contradictory phenomena are
originated from different mechanisms. The investigation of such model, in
our opinion, may lead to several interesting phenomena in understanding
decoherence in quantum open system. We hope that the simple analytical model
described in this paper will assist us to gain further insight in QIP.

\section{ACKNOWLEDGMENTS}

W. L. You is supported by start-up funding at Suzhou University under Grant
No. Q4108907. Y. L. Dong acknowledges the support of the National Natural
Science Foundation of China under Grant No. 10774108 and start-up funding at
Suzhou University under Grant No. Q4108908.


\bigskip

\begin{thebibliography}{99}
\bibitem{ben} C. H. Bennett and S. J. Wiesner, \textrm{Phys. Rev. Lett.}
\textbf{69}, 2881 (1992)

\bibitem{ben92} C. H. Bennett, G. Brassard, C. Crepeau, R. Jozsa, A. Peres,
and W. K. Wootters, \textrm{Phys. Rev. Lett.} \textbf{70}, 1895 (1993)

\bibitem{joz} R. Jozsa and N. Linden,\textrm{\ Proc. R. Soc. Lond. A}
\textbf{459}, 2011 (2003) %quant-ph/0201143.

\bibitem{duan03} L.-M. Duan, E. Demler, and M. D. Lukin, \textrm{Phys. Rev.
Lett.} \textbf{91}, 090402 (2003)

\bibitem{har07} Michael J. Hartmann, Fernando G. S. L. Brand\~{a}o, and
Martin B. Plenio, \textrm{Phys. Rev. Lett.} \textbf{99}, 160501 (2007)

\bibitem{os02} T. J. Osborne, M. A. Nielsen, \textrm{Phys. Rev. A} \textbf{66%
}, 032110 (2002)

\bibitem{ost02} A. Osterloh, L. Amico, G. Falci, and R. Fazio, \textrm{Nature%
} (London) \textbf{416}, 608 (2002)

\bibitem{gun01} D. Gunlycke, V. M. Kendon, V. Vedral, and S. Bose, \textrm{%
Phys. Rev. A} \textbf{64}, 042302 (2001)

\bibitem{ar01} M. C. Arnesen, S. Bose and V. Vedral, \textrm{Phys. Rev. Lett.%
} \textbf{87}, 017901 (2001)

\bibitem{zhu} C. L. Zhang, S. Q. Zhu, and J. Ren, \textrm{Phys. Lett. A}
\textbf{373} 3522 (2009)

\bibitem{wang} X. G. Wang, \textrm{Phys. Rev. A} \textbf{64}, 012313 (2001)

\bibitem{Connor} K. M. O'Connor and W. K. Wootters, \textrm{Phys. Rev. A}
\textbf{63}, 052302 (2001)

\bibitem{Gu} S. J. Gu, S. S. Deng, Y. Q. Li, and H. Q. Lin, \textrm{Phys.
Rev. Lett.} \textbf{93} 086402 (2004)

\bibitem{quan} H. T. Quan, Z. Song, X. F. Liu, P. Zanardi, and C. P. Sun,
\textrm{Phys. Rev. Lett.} \textbf{96}, 140604 (2006)

\bibitem{oru} R. Or\'{u}s, \textrm{Phys. Rev. Lett.} \textbf{100}, 130502
(2008)

\bibitem{amico} L. Amico, R. Fazio, A. Osterloh, and V. Vedral, \textrm{Rev.
Mod. Phys.} \textbf{80}, 517 (2008)

\bibitem{ple} M. B. Plenio and S. F. Huelga, \textrm{Phys. Rev. Lett.}
\textbf{88}, 197901 (2002)

\bibitem{bos} S. Bose, P. L. Knight, M. B. Plenio, and V. Vedral, \textrm{%
Phys. Rev. Lett.} \textbf{83}, 5158 (1999)

\bibitem{yi} X. X. Yi, H. T. Cui, and L. C. Wang, \textrm{Phys. Rev. A}
\textbf{74}, 054102 (2006)

\bibitem{sun} Q. Ai, T. Shi, G. L. Long, and C. P. Sun, \textrm{Phys. Rev. A}
\textbf{78}, 022327 (2008)

\bibitem{zur} W. H. Zurek, \textrm{Phys. Today} \textbf{44}(10), 36 (1991)

\bibitem{yu} T. Yu and J. H. Eberly, \textrm{Science} \textbf{323}, 598
(2009)

\bibitem{bose04} A. Hutton and S. Bose, \textrm{Phys. Rev. A} \textbf{69},
042312 (2004)

\bibitem{zur05} F. M. Cucchietti, J. P. Paz, and W. H. Zurek, \textrm{Phys.
Rev. A} \textbf{72}, 052113 (2005)

\bibitem{wenlong} W.-L You and W.-L. Lu, \textrm{Phys. Lett. A} \textbf{373}%
, 1444 (2009)

\bibitem{sun07} Z. Sun, X. Wang, and C. P. Sun, \textrm{Phys. Rev. A}
\textbf{75}, 062312 (2007)

\bibitem{cheng} W. W. Cheng and J. M. Liu, \textrm{Phys. Rev. A} \textbf{79}%
, 052320 (2009)

\bibitem{ma} X. S. Ma, A. M. Wang, and Y. Cao, \textrm{Phys. Rev. B} \textbf{%
76}, 155327 (2007)

\bibitem{ma08} X. S. Ma, H. S. Cong, J. Y. Zhang, and A. M. Wang, \textrm{%
Eur. Phys. J. D}, \textbf{48}, 285 (2008)

\bibitem{Lieb-AnnPhys-1961} E. Lieb, T. Schultz, and D. Mattis, \textrm{Ann.
Phys.} \textbf{16}, 407 (1961)

\bibitem{Dzyaloshinsky} I. Dzyaloshinsky, \textrm{J. Phys. Chem. Solids}
\textbf{4}, 241 (1958)

\bibitem{Moriya} T. Moriya, \textrm{Phys. Rev. Lett.} \textbf{4}, 228 (1960)

\bibitem{Jordan-Zphys-1928} P. Jordan and E. Wigner, \textrm{Z. Phys.}
\textbf{47}, 631 (1928)

\bibitem{Sachdev} S. Sachdev, \textit{Quantum Phase Transitions}, (Cambridge
University Press, Cambridge, UK, 2000)

\bibitem{yuan07} Z. G. Yuan, P. Zhang, and S. S. Li, \textrm{Phys. Rev. A}
\textbf{76}, 042118 (2007)

\bibitem{Wootters} W. K. Wootters, \textrm{Phys. Rev. Lett.} \textbf{80},
2245 (1998)

\bibitem{sjgu1} S. J. Gu, H. Q. Lin, and Y. Q. Li, \textrm{Phys. Rev. A}
\textbf{68}, 042330 (2003)

\bibitem{sjgu2} S. J. Gu, G. S. Tian, H. Q. Lin, \textrm{Phys. Rev. A}
\textbf{71}, 052322 (2005)

\bibitem{kar} M. Kargarian, R. Jafari, and A. Langari, \textrm{Phys. Rev. A}
\textbf{79}, 042319 (2009)

\bibitem{Cucchietti1} F. M. Cucchietti, J. P. Paz, and W. H. Zurek, \textrm{%
Phys. Rev. A} \textbf{72}, 052113 (2005)

\bibitem{Cucchietti2} F. M. Cucchietti, S. Fernandez-Vidal, and J. P. Paz,
\textrm{Phys. Rev. A} \textbf{75}, 032337 (2007)
\end{thebibliography}
\end{document}